\documentclass[lettersize,journal]{IEEEtran}
\usepackage{amsmath,amsfonts}
\usepackage{algorithmic}
\usepackage{algorithm}
\usepackage{array}
\usepackage[caption=false,font=normalsize,labelfont=sf,textfont=sf]{subfig}
\usepackage{textcomp}
\usepackage{stfloats}
\usepackage{url}
\usepackage{verbatim}
\usepackage{graphicx}
\usepackage{cite}
\usepackage[table]{xcolor}
\usepackage{makecell}
\usepackage{adjustbox}
\usepackage{multirow}
\usepackage{hhline}
\usepackage{siunitx}
\DeclareSIUnit \voltampere { VA }

\begin{document}

\title{An Admittance-Based Inverter Connection Screening Tool for Small-Signal System Strength}

\author{Andreas Hadjileonidas,~\IEEEmembership{Graduate Student Member,~IEEE}, Debargha Brahma,~\IEEEmembership{Member,~IEEE}, Yue Zhu,~\IEEEmembership{Member,~IEEE},   Timothy C. Green,~\IEEEmembership{Fellow,~IEEE}

\thanks{}}

\markboth{\NoCaseChange{\fontsize{7}{8}\selectfont This work has been submitted to the IEEE for possible publication. Copyright may be transferred without notice, after which this version may no longer be accessible.}}%
 {}

\IEEEpubid{}

\maketitle

\begin{abstract}

The increasing occurrence of small-signal instability, particularly sub-synchronous oscillations (SSOs), in power systems with a high penetration of inverter-based resources (IBRs) has made the planning of new IBR connections increasingly important and challenging. 
The impact of such connections on small-signal stability is not always straightforward, as it strongly depends on the connection location, inverter operating mode, control configuration, parametrisation, and operating conditions. 
This paper proposes an inverter connection screening tool (ICST) that enables efficient and accurate assessment of the impact of prospective inverter configurations on small-signal system strength. 
It can identify, among the candidates considered, the most suitable inverter configuration for a given connection location that avoids degrading small-signal system strength and can also enhance it.
As a result, higher IBR penetration can be supported while maintaining small-signal stability. 
The ICST evaluates candidate inverter configurations using their admittances at critical modal frequencies, along with the system's admittance spectrum, thereby avoiding the need for analytical models. 
The ICST-based planning procedure, which can support system operators, asset owners, and IBR developers in decision-making across different stages of planning studies, is demonstrated using a modified IEEE 57-bus system. 
Comparisons with model-based studies demonstrate the accuracy of the ICST in predicting the modal impact of inverter connections and its effectiveness in selecting suitable inverter control configurations.

\end{abstract}


 \section{Introduction}
\IEEEPARstart{I}{nverter} based resources (IBR) are displacing fossil fuel based synchronous generators (SGs) in many systems\cite{Achieving100}. 
Initially, the majority of IBRs were grid-following (GFL) inverters \cite{PowerSystemStabilityIBR}, which posed no significant concerns while power systems were dominated by SGs.
However, as IBR penetration increased and began to dominate the generation mix, % \cite{GPSTagenda}
system strength declined \cite{high_ibr_SOS}, giving rise to new forms of dynamic instabilities \cite{Hatziargyriou2021}. 
Instabilities, especially, sub-synchronous oscillations (SSOs) have repeatedly appeared in IBR-dominated power systems \cite{esig2024oscillations, real_world}, through interactions among IBRs and between IBRs and other components\cite{SSOdeeperPerspective}, such as 
the \qty{8}{\hertz} voltage oscillations in northern Scotland \cite{NESO2024SSO} and the intermittent voltage oscillations in the \qtyrange{17}{20}{\hertz} range in Australia's West Murray area \cite{AEMO2025SSO}. 
This highlights the need to incorporate small-signal strength assessment into IBR planning and connection studies to identify unstable conditions and guide the selection of inverter types and control configurations. 

Traditionally, new connections to the system have been assessed using the short-circuit ratio (SCR) as a system strength metric \cite{scr_ranges}. 
However, SCR, as well as its major variants such as the equivalent short-circuit ratio (ESCR) \cite{ESCR}, assess IBR connections based on the system impedance at only the fundamental frequency \cite{Shah2024}, and therefore do not have a direct connection to the modes of SSOs. 
Recently introduced system strength metrics, such as IMR \cite{IMR} and AM \cite{AM}, bridge this gap by defining \textit{small-signal system strength} as the spatial sensitivity of oscillatory modes.
Nonetheless, assessments of prospective inverter connections remain challenging due to the wide range of available inverter types, control configurations and parametrisations.
Among the types, grid-forming (GFM) inverters \cite{Rosso2021} have attracted significant interest from system operators (SOs) as a way to mitigate system strength decline and avoid SSO instabilities \cite{gfm_bess_damping, high_ibr_SOS}. 
However, interactions among GFM inverters and between GFM inverters and the rest of the network remain understudied\cite{gfm_resonances} and there is scepticism as to whether GFM inverters provide a universal solution to SSOs or may even introduce additional stability challenges \cite{1_hz_GFM}. 
Efficient and effective frameworks are therefore needed for assessing the impact of prospective inverter connections on small-signal system strength. 

In principle, detailed whole-system dynamic models can accurately capture the impact of new IBR connections on small-signal system strength \cite{IMR}, but this approach is particularly challenging. 
First, vendors are reluctant to disclose proprietary high-fidelity models of prospective inverters \cite{Wu2025, IMR}.
Second, the circuit configuration, control configuration, and parametrisation of the prospective inverter may not be known or finalised during the planning stage.
Third, system strength is strongly dependent on operating conditions \cite{am_paper_1}, so system strength assessments must be performed over thousands of operating points \cite{suitable_classification}. 
The number of studies required to cover the combinations of prospective inverter designs and operating points can therefore make detailed model-based assessments computationally impractical \cite{NERC_IBR_LowSCR_2017}.  

To address these challenges, this paper proposes an early-stage inverter connection screening tool (ICST) capable of evaluating the impact of a variety of inverter configurations on small-signal system strength, while accounting for operating point variability and achieving computational efficiency.
The ICST: (a) uses impedance-based whole-system modelling, enabling the analysis of interactions between network components at modal frequencies, (b) requires only IBR admittances at modal frequencies rather than time-domain or state-space models, (c) is computationally efficient, (d) explicitly considers multiple critical operating points, and (e) can be applied during early-stage planning studies for prospective IBRs. 
The ICST results can guide the selection of inverter configurations that avoid destabilising existing poorly damped oscillatory modes and may point to improvements in stability margins, enabling higher IBR uptake while maintaining a stable system. 
The application of the ICST is demonstrated on the IEEE 57-bus system, and the results provide insights into the mode- and location-dependent impact of GFL and GFM inverter connections on system strength.

This paper is structured as follows. Section \ref{sec:preliminary_studies} presents inverter connection case studies that motivate small-signal strength assessment for new IBR connections.
Section \ref{sec:ScreeningTool} presents the ICST, including its underlying methodology, functionality and accuracy assessment, while Section \ref{sec:screening_casestudy} demonstrates its effectiveness through case studies on a modified IBR-dominated IEEE 57-bus system, and discusses its potential use cases.
Finally, Section \ref{conclusions} concludes the paper.
  \section{System Strength and IBR Connections}\label{sec:preliminary_studies} 
\subsection{Small-Signal System Strength} 
This paper adopts a previously introduced spatial small-signal strength metric, Admittance Margin (AM) \cite{am_paper_1}, for new IBR connection assessment. 
AM is defined separately for each oscillatory mode $\lambda=\sigma\pm j\omega$, and each system bus-\emph{k} such that: 
\begin{equation}
    \text{AM}^k_{\lambda} = \frac{|\sigma|}{\|\text{p}_{\text{Y}_\text{Ak}, \lambda}\| } = \frac{|\sigma|}{\|\text{Res}_{\lambda}^* \text{Z}_{kk}^\text{sys}\| },
   \label{eq:am}
\end{equation}
where $|\sigma|$ is the mode margin, i.e., the mode distance to the imaginary axis on the complex plane. $\text{p}_{\text{Y}_{\text{Ak},\lambda}}$ is the participation factor of the apparatus admittance at bus-$k$, $\text{Y}_\text{Ak}$, to $\lambda$, given by the residue of the $k$-th diagonal element of $\text{Z}^\text{sys}(s)$ to $\lambda$, i.e., $\text{Res}_{\lambda}^* \text{Z}_{kk}^\text{sys}$ \cite{grey_box}. $\|\cdot\|$ denotes the Frobenius norm. 

As a complement to the relevant small-signal system strength metric, IMR \cite{IMR}, AM is particularly useful for assessing buses with no connected apparatus, since the corresponding admittance $\text{Y}_\text{Ak}$ can be simply represented as zero. 
Such buses are hereafter referred to as ``empty buses''. 
\subsection{System Strength Assessment - Case study review}\label{sub:case_review}
\begin{figure}[t]
    \centering
    \includegraphics[width=0.85\linewidth]{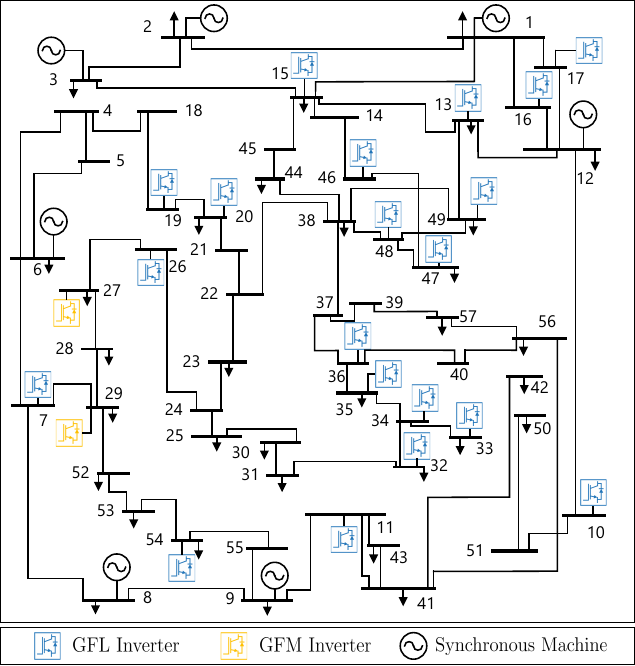}
    \caption{Modified IEEE 57 Bus system   \cite{57BusSystem}}
    \label{fig:57Bus}
\end{figure}

An AM-based small-signal strength assessment was performed in \cite{am_paper_1}.
This paper uses the same test system, a modified IEEE 57-bus system with 22 additional IBRs, shown in \figurename \ref{fig:57Bus}, for IBR connection studies.
The additional IBRs include multiple GFL and GFM inverter configurations and parametrisations to resemble a real-world IBR-dominated grid.
\begin{table}[t] % 57 Bus - v8 results 
        \caption{Oscillatory modes of interest }
    \centering
    \begin{tabular}{|c|c|c|c|} \hline \hline
         Index&  Mode (Hz) &$\zeta (\%)$& Dominant Participants\\ \hline \hline
         $\lambda_1$& $-2.33 + 41.41j$ & 5.6& GFL 19 \& 20\\ \hline 
        $\lambda_2$& $ -2.41 + 28.11j$ &8.6& GFL 32 \& 33\\ \hline 
$\lambda_3$&$ -2.00 + 17.71j $&11.1&GFM 27 \& 29 \\ \hline\hline
    \end{tabular}
    \label{tab:modes}
\end{table}
The AM assessment results highlighted the importance of evaluating small-signal system strength across a set of poorly damped oscillatory modes.
To select the most poorly damped modes, hereafter referred to as ``modes of interest", a $15\%$ damping ratio threshold is used, i.e., 
\begin{equation}
    \zeta=\frac{-\sigma}{\sqrt{\sigma^2 + \omega^2}}\leq15\%. \label{eq:damping_ratio}
\end{equation}
These modes are denoted by $\lambda_i \text{ where } i\in\{1\dots \mathcal{N}\}$, and $\mathcal{N}$ is the number of modes of interest. 
For this test system, three such modes are identified and listed in Table \ref{tab:modes}, together with their damping ratios and dominant IBR participants. 
As revealed by their largest participants, $\lambda_1$ and $\lambda_2$ arise from control interactions between GFL inverters, while $\lambda_3$ from interactions between GFM inverters.
 \begin{table}[t] % 57 Bus - v8 results 
    \caption{AM at empty buses for oscillatory modes of interest }
    \centering
    {\renewcommand{\arraystretch}{0.95}
    \begin{tabular}{|c|c|c|c|}
      \hline  \hline
        Bus&  $\text{AM}_{\lambda_1}$ &  $\text{AM}_{\lambda_2}$  &  $\text{AM}_{\lambda_3}$ \\ 
        \hline\hline
18&\cellcolor[HTML]{00abff}1.16& 18.64& 71.32
\\ \hline  
25& 22.03&\cellcolor[HTML]{00abff}0.75& 86.97
\\ \hline  
52& 95.29& 173.69& \cellcolor[HTML]{00abff}5.63
\\\hline \hline 
    \end{tabular}  
    }
    \label{tab:am_data}
\end{table}
The AM values for all modes of interest are then calculated at the empty buses. 
Three empty buses are chosen for further analysis, each with a low AM for a one of the modes: bus-18 for $\lambda_1$, bus-25 for $\lambda_2$ and bus-52 for $\lambda_3$.
Their corresponding AM values are reported in Table \ref{tab:am_data}, showing that the most critical mode, indicated by the lowest AM in each row, differs across the three buses.
Inverter connection studies at these three buses are presented in subsequent sections of this paper.  
 \subsection{Critical Operating Points Identification}\label{sub:cop}
 \begin{table*}[t]
    \centering
    \caption{COPs identified for bus-25 across system demand ($\%$ of median demand) and IBR generation share scenarios\cite{am_paper_1} }
    \includegraphics[width=.95\linewidth, trim={0cm 0.1cm 0cm .1cm},
    clip]{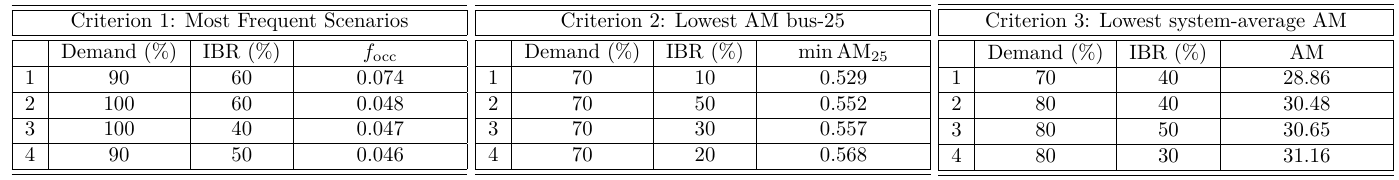}
    \label{fig:COP_25}
     \vspace{-0.2cm}
\end{table*}
Small-signal system strength depends strongly on operating conditions and can vary significantly across operating points. 
Since AM is a small-signal system strength metric, derived from a system linearised around an operating point, it is inherently operating-point dependent and should be recalculated whenever system conditions change. 
However, for large-scale systems with thousands of components, the number of distinct operating points can be enormous, making exhaustive small-signal system strength assessments computationally impractical.
Initial AM studies can be used to identify a representative subset of critical operating points (COPs), reducing the number of cases considered in subsequent studies, while capturing the most critical ones. 
The methodology for identifying COPs is detailed in \cite{am_paper_1}, and as an example, the selected COPs for bus-25 are shown in Table \ref{fig:COP_25}.
Index \emph{c} denotes a critical operating point such that $\emph{c} \in \{ 1,\dots,\emph{C}\}$, with \emph{C} denoting the total number of selected COPs. 
\subsection{Case Study: Inverter Connections System Strength Impact }\label{sub:inverter_connections}
To investigate the impact of different inverter types on the modes of interest, one GFL and one GFM were connected to each of the three buses chosen in Table \ref{tab:am_data}. 
For each inverter connection, $\text{Z}^\text{sys}(s)$ was re-calculated, yielding the updated mode positions, $\lambda_i^\text{final}$. 
The modal shift caused by each connection is defined as $\Delta\lambda_i = \lambda_i^\text{final} - \lambda_i^\text{initial}$. 
The shift in damping for each mode $\Delta\sigma_i = \operatorname{Re}\{\Delta\lambda_i\}$, together with its damping ratio change, $\Delta\zeta_i$, are recorded in Tables \ref{tab:GFL_Effects_withAM} and \ref{tab:GFM_Effects_withAM} for the GFL and GFM connections, respectively. 
\begin{table*}[t]% 57 Bus - v8 results 
   \caption{GFL inverter connection effects on oscillatory modes of interest with the initial AM of the connection buses  }
    \centering
    \begin{tabular}{||c||c|c|c||c|c|c||c|c|c||} \hline \hline
         \cellcolor[HTML]{b0d0eb}GFL
         &\multicolumn{3}{|c||}{Mode $\lambda_1$ - (GFL 19 \& 20)}
         &\multicolumn{3}{|c||}{Mode $\lambda_2$ - (GFL 32 \& 33)}
         &\multicolumn{3}{|c|}{Mode $\lambda_3$ - (GFM 27 \& 29)}\\ \hline \hline
         Bus& $\text{AM}_{\lambda_1}$
         &$\Delta \sigma_1$&  $\Delta \zeta_1(\%)$&   $\text{AM}_{\lambda_2}$&$\Delta \sigma_2$&  $\Delta \zeta_2(\%)$&   $\text{AM}_{\lambda_3}$&$\Delta \sigma_3$&$\Delta \zeta_3(\%)$\\ \hline 
         18& 1.16&0.605
&  -1.3&            18.64 &0.019
&  -0.1&           71.32&-0.003&0.0\\ \hline 
         25& 22.03 &0.031
&  -0.1&  0.75 &1.277
&  -4.1&   86.97&-0.001&0.0\\ \hline 
         52& 95.29 &-0.005
&  0.0&   173.69 &0.000
&  0.0&   5.63&-0.030&+0.1\\\hline\hline
    \end{tabular}
    \label{tab:GFL_Effects_withAM}
\end{table*}
\begin{table*}[t] % 57 Bus - v5 results 
   \caption{GFM inverter connection effects on oscillatory modes of interest with the initial AM of the connection buses }
    \centering
    \begin{tabular}{||c||c|c|c||c|c|c||c|c|c||} \hline \hline
         \cellcolor[HTML]{f7e1b2}GFM
         &\multicolumn{3}{|c||}{Mode $\lambda_1$ - (GFL 19 \& 20)}
         &\multicolumn{3}{|c||}{Mode $\lambda_2$ - (GFL 32 \& 33)}
         &\multicolumn{3}{|c|}{Mode $\lambda_3$ - (GFM 27 \& 29)}\\ \hline \hline
         Bus& $\text{AM}_{\lambda_1}$
         &$\Delta \sigma_1$&  $\Delta \zeta_1(\%)$&   $\text{AM}_{\lambda_2}$&$\Delta \sigma_2$&  $\Delta \zeta_2(\%)$&   $\text{AM}_{\lambda_3}$&$\Delta \sigma_3$&$\Delta \zeta_3(\%)$\\ \hline 
         18& 1.16
        &-1.659
        &  +3.3&            18.64 
        &-0.50&  +0.1&            71.32
        &0.031&-0.2\\ \hline 
                 25& 22.03 
        &0.007&  0.0&   0.75 
        &-1.912&  +5.2&   86.97
        &0.101&-0.5\\ \hline 
         52& 95.29 &0.009&  0.0&   173.69 &-0.009&  0.0&   5.63&2.315&-10.9\\\hline\hline
    \end{tabular}
    \label{tab:GFM_Effects_withAM}
\end{table*}
For all cases in Tables \ref{tab:GFL_Effects_withAM} and \ref{tab:GFM_Effects_withAM}, the largest mode shifts and damping-ratio changes, in either direction, occur for connections at the buses with the lowest AM for each mode. 
Specifically, for $\lambda_1$, inverter connections of both types at bus-18 have a significant impact on the mode, while connections in the other two higher-$\text{AM}$ buses have considerably smaller effects. 
Additionally, the GFL connection at bus-18 significantly destabilises $\lambda_1$, which is expected, given that $\lambda_1$ results from interactions between two GFL inverters at buses 19 and 20 as shown in Table \ref{tab:modes}. 
In contrast, a GFM connection at bus-18 substantially improves the damping of $\lambda_1$. 
A particularly notable case is the GFM inverter connection at bus-52, a low-AM bus for $\lambda_3$ as shown in Table \ref{tab:am_data}.
The mode $\lambda_3$ results from interactions between two GFM inverters at buses 27 and 29. 
It is evident that an additional GFM connection destabilises further this mode, shifting it towards the right-half plane and reducing its damping ratio by $10.9\%$. 
This observation demonstrates that GFM connections do not always enhance system strength, especially for modes that arise from interactions among existing GFM inverters. 
In contrast, Table \ref{tab:GFL_Effects_withAM} shows that a GFL inverter connection at bus-52 shifts $\lambda_3$ slightly left and increases its damping ratio by $0.1\%$, indicating that GFL connections are not always detrimental to system strength. 

The results in Tables \ref{tab:GFL_Effects_withAM} and \ref{tab:GFM_Effects_withAM} show that, despite that low AM buses are considered weak and the corresponding modes are highly sensitive to changes at these buses, they present significant opportunities for system strength improvements, if the appropriate inverter type is connected. 
As discussed, connecting an inverter with characteristics similar to those of the dominant participants of a given mode can further destabilise the mode, whereas a different control configuration may stabilise it. 
However, this strategy relies on detailed knowledge of the critical modes' characteristics and of the control configurations of their dominant inverter participants, information that is typically not available in practice.

Consequently, the most suitable inverter type for a connection location cannot be reliably determined  without explicitly evaluating the impact of each candidate design.
This evaluation process requires multiple designs to be assessed across the COPs, which is computationally intensive and relies on detailed inverter models that are rarely available to SOs.  
These challenges motivate a computationally efficient planning-stage tool for assessing the impact of candidate inverter configurations without requiring their detailed models.

 \section{Inverter connection screening tool}\label{sec:ScreeningTool}
While AM quantifies the participation of a system bus in a given oscillatory mode, it does not directly indicate how an inverter connection will affect that mode.
This impact depends on the mode's underlying characteristics and the connecting inverter configuration.
As demonstrated in Section \ref{sub:inverter_connections}, selecting an appropriate inverter configuration for a given connection location can improve system strength.

Currently, small-signal system strength assessments are largely absent from the planning stage of new IBR connections. 
Although NESO, the SO of Great Britain, recently introduced a set of small-signal assessments as part of the compliance process for new connections \cite{neso_oscillations}, these studies come later in the connection process.
Planning-stage assessments would provide earlier insights into the potential impacts of IBR connections, allowing sufficient time to refine candidate designs and avoid negative effects on system strength and, where possible, support its improvement. 
Additionally, the NESO assessments rely on a Thevenin-equivalent representation of the AC grid at the point of interconnection \cite{neso_oscillations}, which limits their ability to capture IBR interactions \cite{high_ibr_SOS}. 

These limitations motivate the development of the inverter connection screening tool (ICST), designed for the early planning stage of new IBR connections. 
The ICST assesses the impact of candidate inverter configurations on small-signal system strength across multiple operating scenarios, while requiring minimal inverter information and substantially reducing computation time relative to detailed system-strength assessments. 
 \subsection{Prediction of Mode Shift}\label{sub:methodology}
The development of the ICST draws on the concepts of whole-system impedance modelling \cite{whole_system} and participation analysis in impedance models\cite{grey_box}. 
These enable the analysis of dynamic interactions between system components using impedance models, which can also be identified online when white-box models are not available. 

The sensitivity of an eigenvalue $\lambda$ to admittance changes at bus-\emph{k}, defined as the admittance participation factor, is given by the residue of the \emph{k}-th diagonal element of $\text{Z}^\text{sys}$ associated with $\lambda$: 
\begin{equation}
    p_{\text{Y}_\text{Ak}, \lambda} = - \text{Res}^*_{\lambda}\text{Z}^\text{sys}_\text{kk} \label{eq:adm_participation_factor}
\end{equation}
An admittance perturbation at bus-\emph{k}, $\Delta \text{Y}_\text{k}(\lambda)$, results in a shift of the oscillatory mode, $\Delta\lambda$. This is computed as the inner product between the admittance participation factor and the admittance perturbation:
\begin{align}
    \Delta\lambda = 	\langle -\text{Res}_{\lambda}^* \text{Z}_{kk}^\text{sys}, \Delta \text{Y}_\text{Ak}(\lambda)  \rangle
    \label{eq:dl_pred}
\end{align}
At empty buses, the connected admittance is zero ($\text{Y}_\text{Ak}(\lambda) = 0$), however, the admittance participation factor remains defined by \eqref{eq:adm_participation_factor}. 
When a new inverter connection is considered at an empty bus, its impact on the modes of interest can be estimated using \eqref{eq:dl_pred}, with the admittance of the new inverter, evaluated at $ s = \lambda$, $\text{Y}_\text{Ak}(\lambda)$, treated as the admittance perturbation $\Delta\text{Y}_\text{Ak}(\lambda)$.
This is subject to the condition that the size of the potential inverter connection is small compared to the rest of the system seen from that bus. 
To formalise this condition, for a given bus-\emph{k}, the admittances of the inverter, $\text{Y}_\text{Ak}$, and the rest of the grid, $\text{Y}_\text{gk}$, at $ s = \lambda$, are considered. 
Here, $\text{Y}_\text{gk}(\lambda)$ is obtained from the inverse of the diagonal element of the pre-connection whole-system impedance matrix $\text{Y}_\text{gk}(\lambda) = (\text{Z}^\text{sys}_\text{kk}(\lambda))^{-1} $. 
 For the new inverter connection to be treated as a small perturbation, such that  \eqref{eq:dl_pred} remains applicable, the following condition is imposed: 
  \begin{align}
     \frac{\|\text{Y}_\text{Ak}(\lambda)\| }{\|\text{Y}_\text{gk}(\lambda)\|} <0.1. 
 \end{align}
Since the result of \eqref{eq:dl_pred} is based on small-signal assumptions, it provides a prediction of the effect of a new inverter connection on $\lambda$ and is denoted in the remainder of this paper as $\Delta\lambda^\text{pred}$, 
\begin{align}
    \Delta\lambda^\text{pred} =\langle -\text{Res}_{\lambda}^* \text{Z}_{kk}^\text{sys}, \text{Y}_\text{Ak}(\lambda) \rangle.
    \label{eq:delta_lambda_pred_final}
\end{align}
\subsection{Inverter Connection Impact on Damping Ratio}\label{sub:damping_ratio}
\begin{figure}[t]
    \centering
    \includegraphics[width=0.8\linewidth]{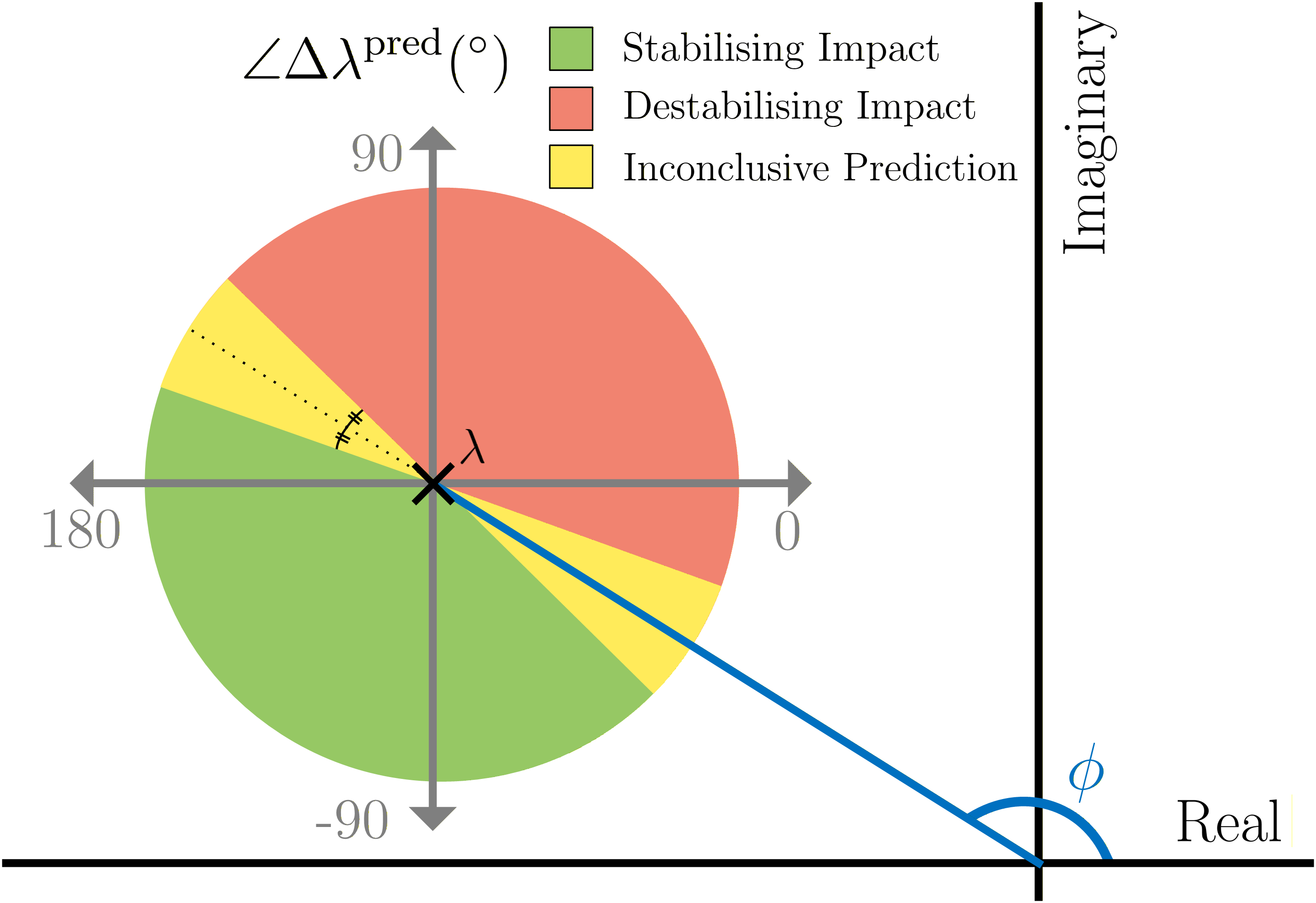}
    \caption{Illustration of the predicted inverter connection impact on mode $\lambda$ }
    \label{fig:illustration_ranges_angles}
\end{figure}

The impact of an inverter connection on $\lambda$, can be assessed from the resulting change in the damping ratio of the mode.
An increase indicates mode stabilisation, whereas a decrease indicates a destabilisation. 
The damping ratio of $\lambda$ can be expressed as $\zeta\! =\!-\text{cos}(\phi)$, where $\phi\! =\! {\scriptstyle\angle}\lambda$, as shown in \figurename \ref{fig:illustration_ranges_angles}. 
Upon the new inverter connection, $\lambda$ will move to a new position, $\lambda^\text{final}\! =\! \lambda + \Delta\lambda$ with a corresponding damping ratio $\zeta^\text{final} = -\text{cos(}\phi^\text{final})$. Hence,
 \begin{align}
     \Delta \zeta = \zeta^\text{final} - \zeta =  \text{cos(}\phi)-\text{cos(}\phi^\text{final}).
 \end{align}
For $\omega\! >\! 0$ and $\sigma\! <\!0$, $\text{cos(}\phi)$ is strictly decreasing, thus for a mode shift $\Delta\lambda^\text{pred} = |\Delta\lambda^\text{pred}| \angle\Delta\lambda^\text{pred}$, the damping ratio changes as follows: 
 \begin{flalign*}
    &\Delta\zeta > 0\text{ if } \phi \leq {\scriptstyle\angle} \Delta\lambda^\text{pred} \leq \phi + 180^\circ \\
   & \Delta\zeta < 0\text{ if } {\scriptstyle\angle}\Delta\lambda^\text{pred}\leq \phi \text{ and }  {\scriptstyle\angle}\Delta\lambda^\text{pred}\geq \phi + 180^\circ, 
\end{flalign*} 
and is illustrated on the complex plane in \figurename \ref{fig:illustration_ranges_angles}.
The final position of $\lambda$ becomes: 
\begin{align}
  \lambda^\text{final}\!=\!\sigma\! + \!|\Delta\lambda^\text{pred}| \cos\!{({\scriptstyle\angle}\Delta\lambda^\text{pred})}\!+\!j (\omega\! +\! |\Delta\lambda^\text{pred}| \sin\!{({\scriptstyle\angle}\Delta\lambda^\text{pred}))} \label{eq:lambda_final}
\end{align}
Combining \eqref{eq:damping_ratio} and  \eqref{eq:lambda_final} yields the expression of $\zeta^\text{final}$ in terms of $\sigma,\omega \text{ and }\Delta\lambda^\text{pred}$ as shown in \eqref{eq:new_damping}. 
The damping ratio change can then be calculated as: $\Delta\zeta =\zeta^\text{final}  - \zeta$.
\begin{figure*}
    \begin{equation}
    \zeta^\text{final} = \frac{-\big(\sigma + |\Delta\lambda^\text{pred}|\cdot \cos{(\angle \Delta\lambda^\text{pred})}\big)}{\sqrt{\sigma^2 +\omega^2 + |\Delta\lambda^\text{pred}|^2 +2\cdot |\Delta\lambda^\text{pred}| \cdot \big(\sigma\cdot \cos{(\angle \Delta\lambda^\text{pred})} + \omega \cdot \sin{(\angle \Delta\lambda^\text{pred})}\bigl) }}
   \label{eq:new_damping}
\end{equation}  
\end{figure*}

\subsection{Most Suitable Inverter Selection}\label{sub:inverter_selection}
An inverter connection affects all oscillatory modes, but to different extents, depending on the admittance participation factor associated with each mode and the inverter admittance.
Therefore, all modes of interest should be considered when assessing the overall impact of a connection on small-signal system strength. 
To account for their relative importance at the connection location, each mode is weighted according to its local criticality, as quantified by its AM.
Specifically, the reciprocal of each mode's AM at the connection bus is used, so that modes with lower AM, corresponding to higher local criticality, are assigned larger weights. 
For a mode $\lambda_i$, its weight at bus-\emph{k}, $ w_{\lambda_i}^k$ is given by:
\begin{align}
    \tilde{w}_{\lambda_i}^k = \frac{1}{\text{AM}^{\text{bus-}k}_{\lambda_i}} \text{ , }          {w}_{\lambda_i}^k = \frac{\tilde{w}_{\lambda_i}^k}{\sum_{i=1}^N \tilde{w}_{\lambda_i}^k}.
    \label{eq:weightings_AM}
\end{align}

\begin{figure}[t]
    \centering
    \hspace{-0.5cm}
    \includegraphics[width=0.9\linewidth]{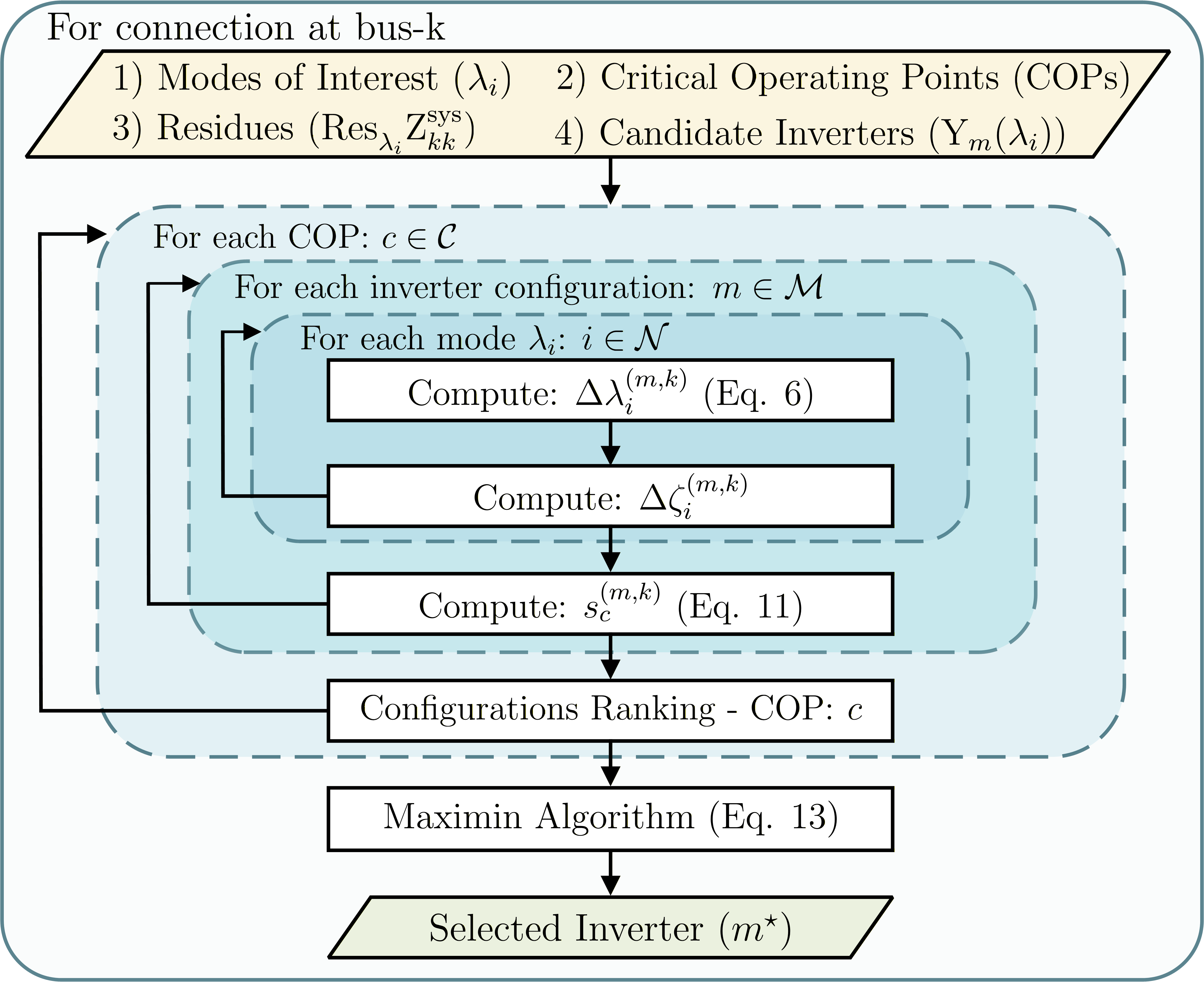}
    \caption{Internal flowchart of the inverter connection screening tool}
    \label{fig:screening_internal}
\end{figure}

%% UPDATED SCORING using damping ratio change. 
Using $\Delta\zeta_i^m$ as a quantitative measure of the impact of an inverter configuration with index $m$ ($m\in \mathcal{M}$, with $\mathcal{M}$ the set of prospective inverter configurations)  on $\lambda_i$ for a COP \emph{c},
the inverter-configuration suitability index is defined:  
 \begin{align}
     s^{(m,k)}_c = \sum_{i=1}^N w_{\lambda_i}^k\cdot \Delta\zeta_i^m,
 \end{align}
to reflect the overall weighted impact of an inverter configuration $m$ connection at bus-\emph{k}, on all modes of interest. 
To select the most suitable inverter configuration across all COPs, priority is given to minimise the connecting inverter's worst-case impact. 
Therefore, a maximin algorithm is employed to guide the selection by identifying configurations with the highest minimum score across all COPs, thereby minimising the connected inverter's worst-case contribution.
\begin{align}
    \underline{s}^{{(m,k)}} = \min_{c\in \mathcal{C}} s^{(m,k)}_c 
    \label{eq:maximin_1}
\end{align}
The selected inverter index $\emph{m}^\star$ is consequently given by: 
\begin{align}
    \emph{m}^\star = \arg \max_{m\in \mathcal{M}}\underline{s}^{{(m,k)}} = \arg \max_{m\in \mathcal{M}}\min_{c\in \mathcal{C}} s^\text{(m,k)}_c
     \label{eq:maximin_2}
\end{align}
A flowchart of the internal ICST processes is shown in \figurename \ref{fig:screening_internal}.

\subsection{Accuracy assessment of prediction method}
A key condition applied in \eqref{eq:dl_pred} is that the admittance perturbation $\Delta \text{Y}_\text{Ak}(\lambda)$ is very small compared to the grid admittance, $ \text{Y}_\text{gk}(\lambda)$. In \eqref{eq:delta_lambda_pred_final}, this perturbation is taken as the admittance of a full-scale inverter $\text{Y}_\text{Ak}(\lambda)$, therefore, to identify the factors influencing the accuracy of $\Delta\lambda^\text{pred}$, we first revisit the derivation of admittance participation factor in \cite{grey_box}.

Let the \emph{k}-th diagonal entry of the whole-system impedance matrix be $G(s)\!=\!\text{Z}^\text{sys}_{kk}(s)$.
$G(s)$ is a $2\!\times\!2$ transfer-function matrix, and define $H(s)\! =\! G^{-1}(s)$. As introduced in \cite{grey_box}, a mode $\lambda$ is a zero of the determinant of $H(s)$. For simplicity, we denote $H^{(0)}_\text{det}\!=\!\operatorname{det}(H^{(0)}\!)$, such that $H_\text{det}^{(0)}(\lambda)\! =\!0$, 
where superscript $(0)$ denotes the system before perturbation. 
After a perturbation in $G$, pole $\lambda$ moves by $\Delta\lambda$ and: 
\begin{align}
    H_\text{det}^{(1)}(\lambda+\Delta\lambda) = 0, \label{eq:det_lambda_delta_lambda}
\end{align}
where superscript $(1)$ denotes the system after perturbation.
Since $H_\text{det}^{(1)}(s)$ is analytic around $\lambda$, we can take its first-order Taylor expansion and neglect higher-order terms: 
\begin{align}
    H_\text{det}^{(1)}(\lambda+\Delta\lambda) = H_\text{det}^{(1)}(\lambda) + H_\text{det}'^{(1)}(\lambda)\cdot \Delta\lambda.
    \label{eq:taylor}
\end{align} 
Combining \eqref{eq:det_lambda_delta_lambda} and \eqref{eq:taylor} yields: 
\begin{align}
    H_\text{det}^{(1)}(\lambda) + H_\text{det}'^{(1)}(\lambda)  \cdot\Delta\lambda =  0.
    \label{eq:taylor_2}
\end{align}
Adding ${H_\text{det}}^{\prime(0)}(\lambda)\cdot\Delta\lambda$ to both sides and defining  $\Delta H_\text{det} = H_\text{det}^{(1)}- H_\text{det}^{(0) }$ gives:
\begin{align}
    \Delta H_\text{det}(\lambda)+ H'^{(0)}_\text{det} (\lambda)\cdot\Delta\lambda + \Delta{H'_\text{det}}(\lambda)\cdot\Delta\lambda = 0 .\label{eq:grey_box_proof_before_suppressing}
\end{align}
We get a first-order approximation by neglecting the high-order term $\Delta{H'_\text{det}}(\lambda)\cdot\Delta\lambda$ (as done in \cite{grey_box}): 
\begin{align}
    \Delta\lambda = - \big({H'}_\text{det}^{(0)}(\lambda)\big)^{-1} \cdot \Delta H_\text{det}(\lambda)= -\frac{\Delta H_\text{det}(\lambda)}{{{H}^{\prime(0) }_\text{det}}(\lambda) },
    \label{eq:delta_labmda}
\end{align}
which provides $\Delta\lambda^\text{pred}$ as described in \eqref{eq:delta_lambda_pred_final}.
If the higher-order term is retained in \eqref{eq:grey_box_proof_before_suppressing}, we have: 
\begin{align}
    [{\Delta H^{\prime}_\text{det}}(\lambda) +{H^{\prime(0)}_\text{det}}(\lambda)] \cdot \Delta\lambda^\text{HO} = -  \Delta H_\text{det}(\lambda),
\end{align}
where $\Delta\lambda^\text{HO}$ refers to the change of $\lambda$ considering the high-order term such that:
\begin{align}
    \Delta\lambda^\text{HO} =  -\frac{ \Delta H_\text{det}(\lambda)}{ \Delta {H'_\text{det}}(\lambda) +  {H'_\text{det}}^{(0)}(\lambda)  } .\label{eq:dl_ho}
\end{align}
Combining \eqref{eq:delta_labmda} and \eqref{eq:dl_ho} and rearranging the equation yields
\begin{align}
    &\frac{\Delta\lambda^\text{pred}}{\Delta\lambda^\text{HO}} =  \frac{ \Delta {H'_\text{det}}(\lambda) +  {H^{\prime(0)}_\text{det}}(\lambda)  }{ {H^{\prime(0)}_\text{det}}(\lambda) } = 1+ \frac{\Delta {H_\text{det}}(\lambda)} { {H^{\prime(0)}_\text{det}}(\lambda) }   \notag\\
    &=  1+ \frac{ {H^{\prime(1)}_\text{det}}(\lambda) - { {H^{\prime(0)}_\text{det}}(\lambda) }}{H^{\prime(0)}_\text{det}(\lambda)} = \frac{{H^{\prime(1)}_\text{det}}(\lambda)}{{H^{\prime(0)}_\text{det}}(\lambda)}.
\end{align}
After a new inverter connection at an empty bus-\emph{k}, the corresponding diagonal element of $\text{Z}^\text{sys}$ is given by: 
\begin{align}
     \text{Z}^\text{sys (1)}_\text{kk} = (H^\text{(1)}(s))^{-1} = (\text{Y}_\text{gk}(s) + \text{Y}_\text{A}(s))^{-1},
\end{align}
which yields,% which yieds: 
\begin{align}
    \frac{\Delta\lambda^\text{pred}}{\Delta\lambda^\text{HO}} = \frac{H_\text{det}^{\prime(1)}(s)|_{s=\lambda}}{ H_\text{det}^{\prime(0)}(s)|_{s=\lambda}} = \frac{\frac{d}{ds}{\text{det}[\text{Y}_\text{gk}(s) + \text{Y}_\text{A}(s)]}|_{s=\lambda}}{\frac{d}{ds}{\text{det}(\text{Y}_\text{gk}}(s))|_{s=\lambda}}
    \label{eq:dl_pred_exact}
\end{align}
Therefore, inaccuracies in $\Delta\lambda^\text{pred}$ arise when \eqref{eq:dl_pred_exact} deviates from unity. 
Applying Jacobi's formula to \eqref{eq:dl_pred_exact} yields: 
\begin{align}
     & \frac{\Delta\lambda^\text{pred}}{\Delta\lambda^\text{HO}} = \frac{\text{Tr}[\text{adj}(\mathrm{Y}_\text{gk}+\mathrm{Y}_\text{A})\cdot \frac{d}{ds}(\mathrm{Y}_\text{gk}+\mathrm{Y}_\text{A})]}{\text{Tr}[\text{adj}(\mathrm{Y}_\text{gk})\cdot \frac{d}{ds}(\mathrm{Y}_\text{gk})]}\notag \\
    & =  1\!  
    +\! \frac{\text{Tr}[\text{adj}(\mathrm{Y}_{\mathrm{gk}}) \!\mathrm{Y}'_\mathrm{A}] 
    +\! \text{Tr}[\text{adj}(\mathrm{Y}_\mathrm{A})   \!\mathrm{Y}'_\mathrm{gk}] 
    +\! \text{Tr}[\text{adj}(\mathrm{Y}_\mathrm{A}) \mathrm{Y}'_\mathrm{A}]}{\text{Tr}[\text{adj}(\mathrm{Y}_\mathrm{gk}) \mathrm{Y}'_\mathrm{gk}]}\label{eq:dl_pred_dl_ho_final}
    \end{align}
where $\text{Y}'_\text{gk}$ and $\text{Y}'_\text{A}$ denote the derivatives of $\text{Y}_\text{gk}$ and $\text{Y}_\text{A}$ to $s$, respectively.
Based on (48) in \cite{grey_box}, the denominator of \eqref{eq:dl_pred_dl_ho_final} is derived:
\begin{align}
    \text{Tr}[\text{adj}(\text{Y}_\text{gk})\cdot \text{Y}_\text{gk}^{\prime}]_{s=\lambda} = \frac{[\text{adj}(\text{Y}_\text{gk}(\lambda))]_{ij}}{[\text{Res}_\lambda \text{Z}^\text{sys}_\text{kk}]_{ij} },
\end{align}
where $(i, j)\in\{d,q\}^2$.
\begin{figure}[t]
    \centering
    \includegraphics[width=1\linewidth]{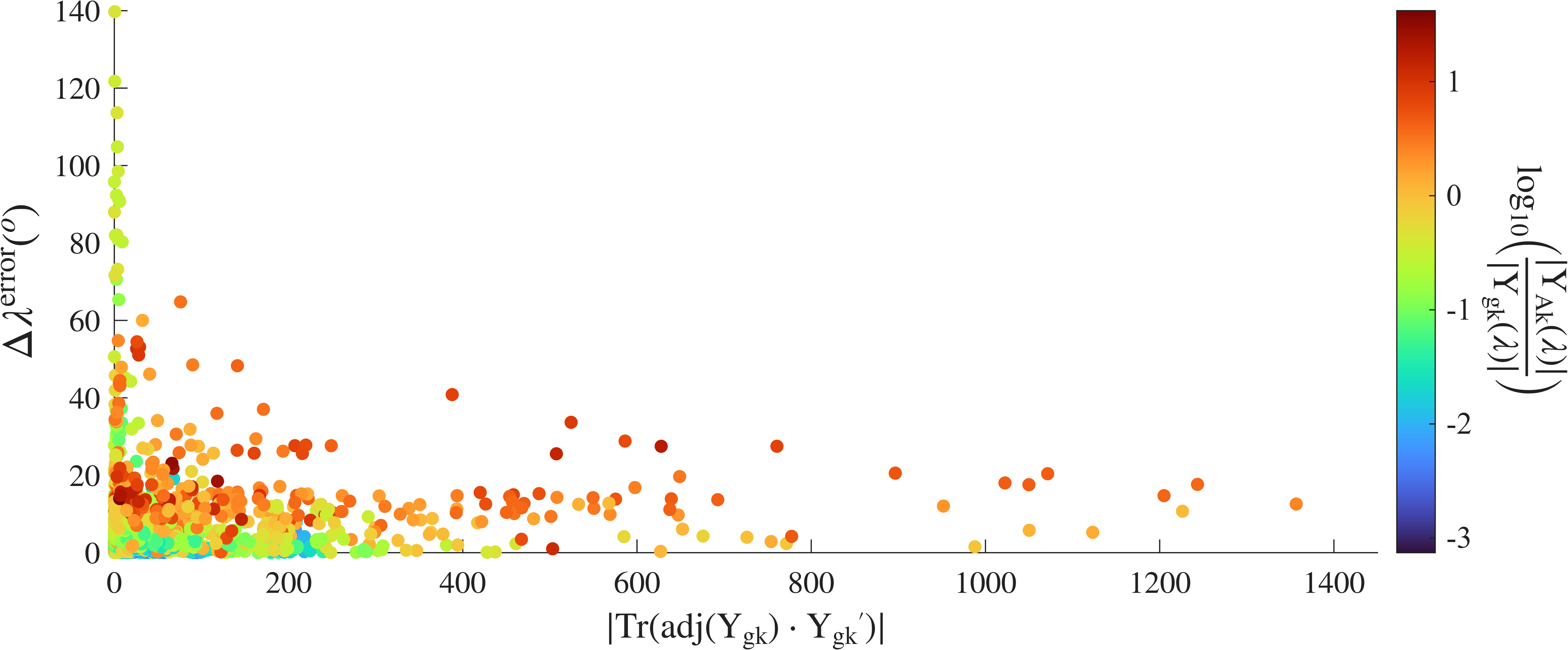}
    \caption{Empirical results of method's accuracy against $|\text{Tr}[\text{adj}(\text{Y}_\text{gk})\cdot \frac{d\text{Y}_\text{gk}}{ds}]_{\text{s=}\lambda}|$ }
    \label{fig:DL_Error_Hdet_der}
     \vspace{-0.3cm}
\end{figure}
The pre-connection whole-system impedance provides the values of  $\text{Res}_\lambda \text{Z}^\text{sys}_\text{kk}$ and $\text{adj}(\text{Y}_\text{gk}(\lambda))$ which can be used to obtain $\text{Tr}[\text{adj}(\text{Y}_\text{gk})\cdot \text{Y}_\text{gk}^{\prime}]_{s=\lambda}$.
As evident from \eqref{eq:dl_pred_dl_ho_final}, larger values of $\text{Tr}[\text{adj}(\text{Y}_\text{gk})\cdot \text{Y}_\mathrm{gk}^{\prime}]_{s=\lambda}$ drive the ratio $\Delta\lambda^\text{pred}/{\Delta\lambda^\text{HO}}$ toward unity, improving the accuracy of $\Delta\lambda^\text{pred}$. 
This is supported by the empirical results in \figurename \ref{fig:DL_Error_Hdet_der}. Specifically, the angle error between $\angle\Delta\lambda^\text{pred}$ and $\angle\Delta\lambda^\text{actual}$ is plotted against $|\text{Tr}[\text{adj}(\text{Y}_\text{gk}) \cdot\! \text{Y}_\text{gk}^{\prime}]_{s=\lambda}|$. 
The results show that all high error cases are associated with either small $|\text{Tr}[\text{adj}(\text{Y}_\text{gk})\cdot\! \text{Y}_\text{gk}^{\prime}]_{s=\lambda}|$ or large $\frac{\|\text{Y}_\text{Ak}(\lambda)\|}{\|\text{Y}_\text{gk}(\lambda)\|}$. The former violates the condition set in this Section, while the latter violates the condition introduced in Section \ref{sub:methodology}.
\subsection{Screening Tool Precision}\label{sub:precision}
\subsubsection{Setup}
To validate the precision of results obtained through \eqref{eq:delta_lambda_pred_final}, $\Delta \lambda^\text{pred}$ is compared to the actual mode shift upon an inverter connection, $\Delta\lambda^\text{actual}$. 
To obtain $\Delta\lambda^\text{actual}$, $\text{Z}^\text{sys}(s)$ is re-calculated post-connection, to give $\Delta\lambda^\text{actual} = \lambda^{(1)} - \lambda^{(0)}$. 
This procedure is performed for the baseline system operating scenario of the test system in Section \ref{sec:preliminary_studies}, and is repeated for a total of 10976 cases (connections of 392 distinct inverter configurations at each of the 28 empty system buses).  
\subsubsection{Prediction Results Comparison}\label{sub:method_accuracy}
\begin{figure}[t]% 57 Bus - v6 results 
    \centering
    \includegraphics[width=1\linewidth]{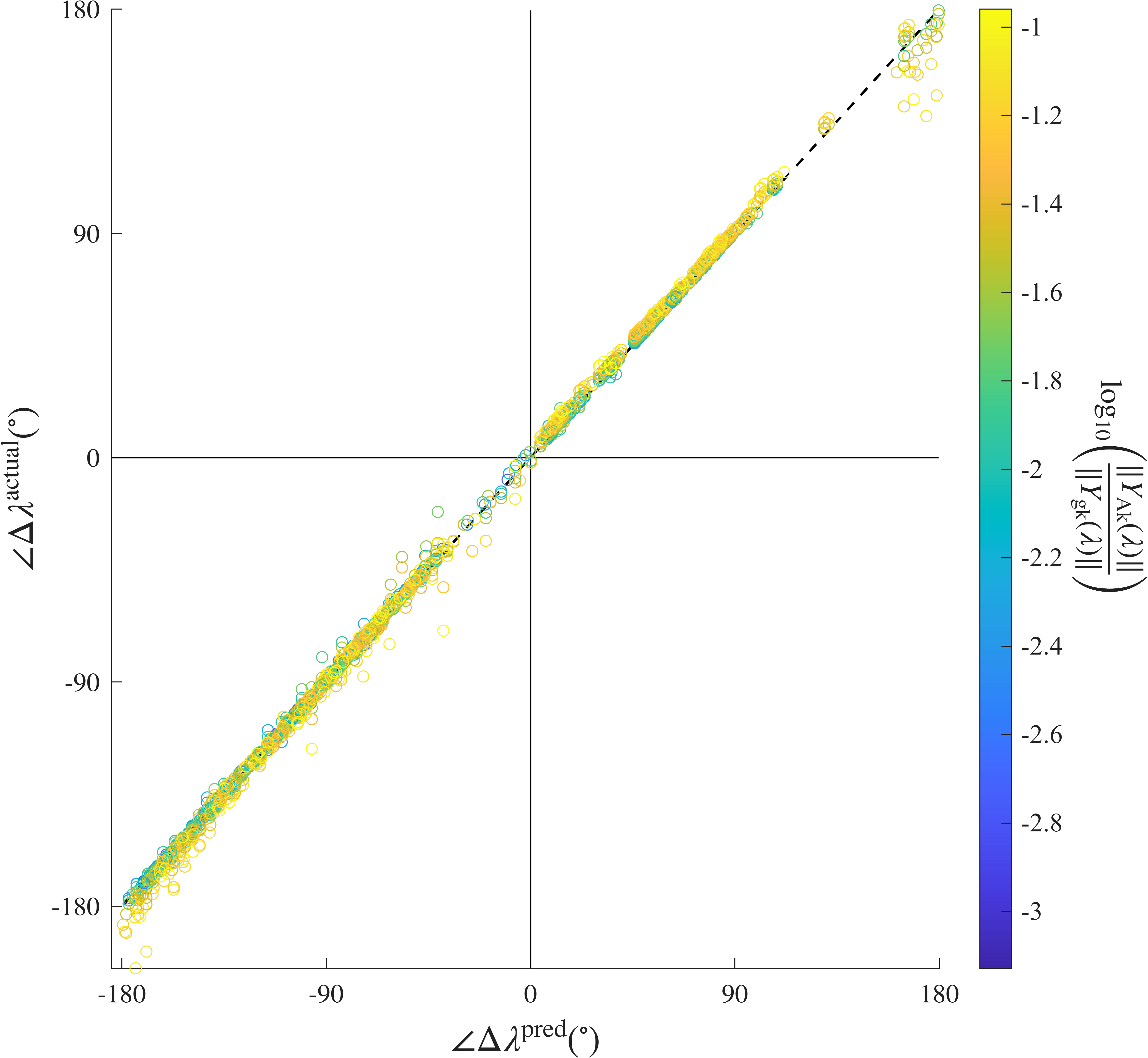}
    \caption{ $\angle\Delta \lambda^\text{actual}$ vs $\angle\Delta\lambda^\text{pred}$. Each data point represents a separate test case. Dotted line is $\angle\Delta\lambda^\text{actual} = \angle\Delta \lambda^\text{pred} $. Cases shown for $\frac{\|\text{Y}_\text{Ak}\|}{\|\text{Y}_\text{gk}\|}<0.1$}
    \label{fig:delta_lambda_angle_comparison}
    \vspace{-0.3cm}
\end{figure}
 \begin{table}[t] % 57 Bus - v6 results 
     \caption{Method Error Statistics for different thresholds of $ {\frac{\|\text{Y}_\text{A}(\lambda)\|}{\|\text{Y}_\text{g} (\lambda)\|} }$}
     \centering
          \begin{tabular}{|c|c|c|c|c|c|c|} \hline
    \multicolumn{2}{|c|}{} & \multicolumn{5}{c|}{\textbf{ 
     $ \frac{\|\text{Y}_\text{Ak}(\lambda)\|}{\|\text{Y}_\text{gk} (\lambda)\|}$  Thresholds}} \\ \hhline{|~~|-----|}
        \multicolumn{2}{|c|}{}  
        & \cellcolor[HTML]{f0f0f0}0.01  & \cellcolor[HTML]{f0f0f0}0.05 & \cellcolor[HTML]{f0f0f0}0.10 & \cellcolor[HTML]{f0f0f0}0.50 & \cellcolor[HTML]{f0f0f0}1 \\ \hline
        \multirow{4}{*}{\rotatebox{90}{$\Delta\lambda^\text{error}$}}
        & Mean $(^\circ)$
        & 0.65  & 1.10 & 1.64 & 3.42 & 3.68 \\ \cline{2-7}
        & Median $(^\circ)$
        & 0.30  & 0.71 & 0.95 & 1.41 & 1.52 \\ \cline{2-7}
        & Q3 (75\%) $(^\circ)$
        & 0.65  & 1.38 & 2.00 & 3.36 & 3.69 \\ \cline{2-7}
        & $95^{\text{th}}$ percentile $(^\circ)$
        & 2.19  & 3.36 & 4.87 & 10.22 & 11.72 \\ \hline
    \end{tabular}
     \label{tab:statistics}
     \vspace{-0.2cm}
 \end{table}
To facilitate comparison,  $\angle \Delta\lambda^\text{actual}$ and $\angle \Delta\lambda^\text{pred}$ are plotted against each other in \figurename \ref{fig:delta_lambda_angle_comparison}, where data points closer to the dashed $\angle \Delta\lambda^\text{actual} =\angle \Delta\lambda^\text{pred}$ line indicate higher prediction precision. 
The results are plotted for $\frac{\|\text{Y}_\text{Ak}(\lambda)\|}{\|\text{Y}_\text{gk}(\lambda)\|} <0.1$, the condition set for this method, and demonstrate a strong correlation between $\angle\Delta\lambda^\text{actual}$ and $\angle \Delta\lambda^\text{pred}$, confirming the high accuracy of the method in predicting the impact of new inverter connections to oscillatory modes.  
Furthermore, Table \ref{tab:statistics} reports the $\Delta\lambda^\text{error}$ statistics for different thresholds of $ {\frac{\|\text{Y}_\text{A}(\lambda)\|}{\|\text{Y}_\text{g} (\lambda)\|} }$, demonstrating that the method's accuracy deteriorates for thresholds
beyond 0.1.

\subsubsection{Incorporating Prediction Error}\label{subsub:prediction_error}
As discussed in Section \ref{sub:inverter_selection}, for $\angle\Delta \lambda^\text{pred}\! \in\! \{\phi, \phi\!+\!180^\circ\!\}$, an inverter connection neither stabilizes nor destabilizes a mode, as its damping ratio remains unchanged. To account for accuracy errors in $\Delta\lambda^\text{pred}$, a buffer band is introduced around these two angles, labelled as ``Inconclusive prediction'' in \figurename \ref{fig:illustration_ranges_angles}.
If $\angle\Delta \lambda^\text{pred}$ falls within one of the buffer bands, the cautious approach is to assume that the inverter has a destabilising impact.
Based on the error statistics in Table \ref{tab:statistics}, and after applying the $ \frac{\|\text{Y}_\text{Ak}\|}{\|\text{Y}_\text{gk}\| } <0.1$ threshold,  the buffer band is empirically set to the 95th-percentile error across all tested cases, resulting in a band of $\pm 4.87^\circ$. 
 To summarise, \figurename \ref{fig:illustration_ranges_angles} illustrates the polar diagram used to classify the calculated $\angle\Delta\lambda^\text{pred}$ values into three regions: 
\begin{itemize}
    \item \textbf{Stabilising region}: The inverter connection improves the oscillatory mode's damping ratio,
    \item \textbf{Destabilising region}: The inverter connection reduces  the oscillatory mode's damping ratio,
    \item \textbf{``Inconclusive Prediction'' region} around the boundaries between stabilising and destabilising regions. 
\end{itemize}
 \vspace{-0.3cm}
\subsection{Screening Tool vs. Full System Recalculation Comparison}
To illustrate the benefits of the ICST, Table \ref{tab:screening_comparison} compares it with eigenvalue-based assessments using the re-computed model-based $\text{Z}^\text{sys}$ that includes the new inverter model. 
This comparison considers the duration, required information, and accuracy of each method. 
The screening tool substantially reduces computation time relative to model-based studies while requiring significantly less information about the inverter and the system.
Although detailed studies provide more accurate predictions of an inverter's impact, the ICST provides fast, binary, and interpretable results, suitable for screening assessments during the new connection planning stage.  
\begin{table*}[t]
\centering
\renewcommand{\arraystretch}{1.25}
       \caption{Comparison between screening tool and detailed studies for 392 distinct inverter configurations}
       \begin{tabular}{|c|c|c|}\hline\hline
         \textbf{Criterion}&    \textbf{Inverter Connection Screening Tool}& \makecell{ \textbf{Detailed Studies}} \\\hline\hline
         \textbf{Computational Time}&  $< 1$ minute for all 392 configurations &$392  \cdot 20$ seconds = 2 hours and 10 minutes * \\\hline
         \textbf{Information Required}&   \makecell{Pre-connection system data ($\text{Res}_\lambda Z^\text{sys}$) and $\mathrm{Y}_\mathrm{A}(\lambda)$}& \makecell{ Full system model \& Detailed vendor inverter models}\\\hline 
        \textbf{ Result Interpretation}&    \makecell{Binary result (Stabilising-Destabilising effect)} &  \makecell{More accurate %but requires detailed analysis
        }\\\hline
\textbf{Purpose} &
\makecell{
Fast screening of multiple inverter configurations } &
\makecell{
Detailed validation of selected cases
} \\\hline
\multicolumn{3}{@{}r@{}}{\footnotesize * Based on Simplus Grid Tool~\cite{simplus} on a standard computer}
\end{tabular}
    \vspace{-0.2cm}
    \label{tab:screening_comparison}
\end{table*}
\section{Screening Tool Applications }\label{sec:screening_casestudy}
The newly developed metric, AM, captures the spatial variation in small-signal system strength and reveals its sensitivity to operating conditions, 
 with high and low values indicating strong and weak buses, respectively. 
\begin{figure*}[!t]
    \centering
    \includegraphics[width=0.93\linewidth]{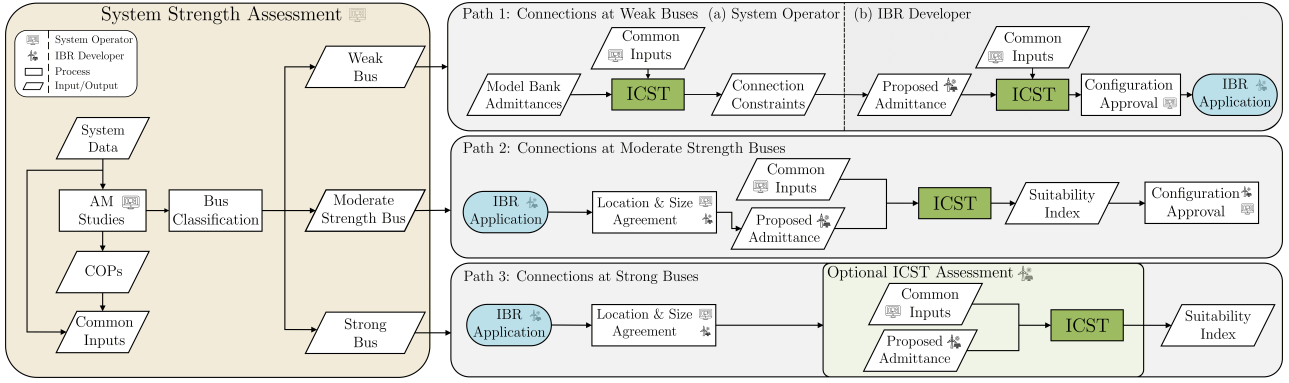}
    \caption{Flowchart demonstrating small-signal strength-specific planning procedure and potential applications of the inverter connection screening tool.}
    \vspace{-0.3cm}
    \label{fig:screening_flowchart}
\end{figure*}
The results in Section \ref{sec:preliminary_studies} demonstrate that weak buses present opportunities for small-signal strength enhancements, provided an appropriate inverter configuration is introduced. In contrast, at strong buses, oscillatory modes are less sensitive to new inverter connections and can therefore be assigned lower priority in small-signal strength assessments. 
Based on these conclusions, three application paths for the ICST are proposed for the early planning stage of new IBR connections, each corresponding to a different level of connection-bus strength.
A flowchart illustrating these application paths is shown in \figurename~\ref{fig:screening_flowchart}.
It is worth noting that selecting the appropriate path for a given bus requires a prior AM-based system strength assessment.

Although the ICST is primarily intended to assess connections at empty buses, it is not limited to such cases. For a bus with existing apparatus, a virtual empty bus can be assumed in close proximity, to enable the application of the ICST. 
\vspace{-0.3cm}
\subsection{Suggested Screening Tool Uses}
This sub-section introduces the three paths in detail.
\subsubsection{Path 1: Connections at Weak Buses}
For weak buses, the SO's approach for new inverter connections should focus on
(a) avoiding further destabilisation and (b) possibly enhancing system stability margins.
Using inverter admittances from a model bank i.e. an SO-developed library of inverter models including different types, configurations and parametrisations, the SO can apply the ICST to screen inverter configurations and identify the most suitable to meet these objectives.
The identified inverter configurations can then be used to define connection constraints for the given connection bus, in terms of inverter type (GFL/GFM), control configuration (e.g. GFL with P-f droop) and parameterisation, or target inverter admittance profiles over specific frequency ranges.
Once connection constraints are established, IBR developers can use the ICST to screen their inverter designs, assess their eligibility against these constraints, and select the most suitable inverter design before submitting an IBR development application. 
This procedure is illustrated in path 1 of the flowchart in \figurename \ref{fig:screening_flowchart}.

\subsubsection{Path 2: Connections at Moderate-Strength Buses}
For buses with moderate strength, the SO may allow developers to submit connection applications without imposing explicit constraints. Interested developers can submit an application for a specific plant location and IBR capacity. 
Once the initial application is approved by the SO, the developer can use the ICST in their design process to select an appropriate inverter design. 
This procedure is illustrated in path 2 in \figurename \ref{fig:screening_flowchart}.

\subsubsection{Path 3: Connections at Strong Buses}
For strong buses, the use of ICST is voluntary, as shown in path 3 of \figurename \ref{fig:screening_flowchart}.
This is supported by the case study in Section \ref{sub:inverter_connections}, which shows that inverter connections at high-AM buses have a limited impact on oscillatory modes due to their low sensitivity to admittance changes at the connection bus. 
\subsection{Inverter Connection Screening Tool - Case Study}
The case study presented in this section demonstrates the primary application of the ICST, namely screening candidate configurations for a prospective inverter connection.
%Table showing mode positions, angle phi and ranges
Building on the system-strength study of the modified IEEE 57-bus system in Section \ref{sub:case_review}, a potential inverter connection at bus-25 is considered.
In this case study, the inverter size and connection location are assumed to be determined. 
For this connection location, the COPs are determined from the AM assessment and listed in Table \ref{fig:COP_25}. 
To illustrate the ICST results, the most frequently occurring operating scenario is selected, corresponding to demand = 90\% and IBR = 60\%.
Table \ref{tab:ranges} shows the AM values of the three modes of interest at bus-25 for this COP, which are used to compute the weights associated with each mode according to \eqref{eq:weightings_AM}.
For this screening study, a candidate set of 392 inverter designs is considered, comprising different control configurations and parametrisations for both GFL and GFM inverters. The five inverter control configurations considered are: 
\begin{itemize}
    \item \textbf{GFL-A}: GFL - no outer loop controllers
    \item \textbf{GFL-B}: GFL - DC voltage control
    \item \textbf{GFL-C}: GFL - DC and AC voltage control
     \item \textbf{GFM-A}: GFM - P-f droop control 
    \item \textbf{GFM-B}: GFM - P-f and Q-$\text{V}_\text{AC}$ droop control
\end{itemize}
The choice of inverter configurations will depend on the manufacturer or developer portfolios and, for SOs, on their available model banks. 
For this study, these five configurations capture a representative set of commonly used inverter control configurations \cite{gfl2022,gfm2021}, however, the candidate set can be extended to include other configurations.
The rated capacity of all inverter configurations is fixed at 0.2 p.u. on a \qty{100}{\mega\voltampere} base. %1.6\% of the median system demand).
For each mode, the admittance participation factor is calculated from $\text{Z}^\text{sys}(s)$ using \eqref{eq:adm_participation_factor}, while the admittance of each of the 392 inverter configurations is evaluated at $s=\lambda_i$, $\text{Y}_\text{A-m}(\lambda_i)$.
\begin{table*}[!t]
    \centering
    \vspace{-0.2cm}
        \caption{Stabilising, destabilising and inconclusive ranges for $\angle\Delta\lambda^\text{pred}$ for the three modes of interest}% at COP: Demand = 90\%, IBR = 60\%}        
        \begin{tabular}{|c|c|c|c||c|c|c|c|}\hline\hline
         &  $\lambda_i \text{ (COP: 90/60)}$ & $\text{AM}_{\lambda_i}^\text{bus-25}$  &$w_{\lambda_i} (\%)$&  $\phi_i$&  \textbf{Stabilising Range}&  \textbf{Destabilising Range}& \textbf{Inconclusive Prediction Ranges}\\\hline\hline
         $\lambda_1$&  -1.98 + 41.34j  & 15.55&4.0&  $92.74^\circ$&  
         $ (97.61^\circ ,267.87^\circ)$&
         % $\angle\Delta\lambda^\text{pred} \in [-82.39^\circ, 87.87^\circ]$& 
         $[-82.39^\circ, 87.87^\circ]$& 
         $(87.87^\circ,97.61^\circ] \cup [-92.13^\circ ,-82.39^\circ) $\\\hline
         $\lambda_2$&  -2.16 + 28.34j & 0.66&94.7&  $94.36^\circ$&
         $(99.23^\circ, 269.49^\circ)$& 
         $[-80.77^\circ, 89.49^\circ]$& 
         $(89.49^\circ, 99.23^\circ] \cup [-90.51^\circ ,-80.77^\circ) $\\\hline
         $\lambda_3$&
         -0.83  + 17.28j & 49.05  &1.3&  $92.75^\circ$&
         $(97.62^\circ, 267.88^\circ)$ &
         $[-82.38^\circ, 87.88^\circ]$&
         $(87.88^\circ, 97.62^\circ] \cup [-92.12^\circ, -82.38^\circ) $\\ \hline\hline
    \end{tabular}
    \label{tab:ranges}
        \vspace{-0.2cm}
\end{table*}
The predicted impact of each inverter configuration on the modes of interest, $\Delta\lambda^\text{pred}_i$, is computed using  \eqref{eq:delta_lambda_pred_final}. 

For the chosen operating scenario, the positions of the three modes of interest are identified and used to determine the angles $\phi_i$ from which the stabilising, destabilising and inconclusive prediction ranges are derived, as outlined in Section \ref{sec:ScreeningTool}.
These ranges are listed in Table \ref{tab:ranges}. 
\begin{figure*}[!t]
    \centering
    \includegraphics[width=.95\linewidth]{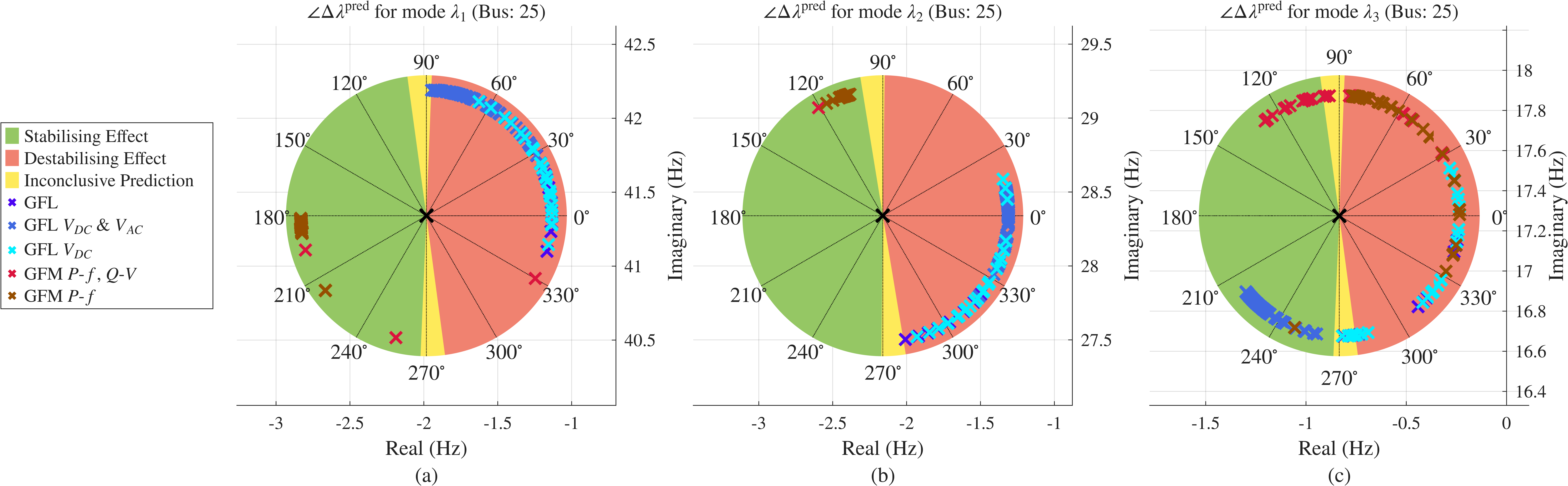}
    \caption{Screening study results for inverter connection at bus-25, shown for the three modes of interest at the COP: Demand = 90\%, IBR = 60\% }
    \label{fig:Screening3Modes}
\end{figure*}
The screening results for all inverter configurations, evaluated separately for each mode of interest, are shown in the polar plots of \figurename  \ref{fig:Screening3Modes}.

From \figurename  \ref{fig:Screening3Modes} (b), for mode $\lambda_2$, every GFM inverter connection improves the mode's damping ratio, whereas every GFL inverter connection has the opposite effect, further destabilising the mode. 
These results show a clear distinction between the two primary inverter types, making the inverter type selection relatively straightforward for this mode. 
The results for mode $\lambda_1$, shown in \figurename  \ref{fig:Screening3Modes} (a), follow a similar trend to $\lambda_2$, as all GFM inverter configurations, except one, have a stabilising impact, while every GFL inverter has a destabilising impact. 

The single destabilising GFM configuration highlights the role of control parametrisation and shows that selecting the inverter type alone is not sufficient to determine connection suitability. 
In the case of mode $\lambda_3$, the results in \figurename  \ref{fig:Screening3Modes} (c)
show no clear separation by inverter type. 
Instead, different control configurations of the same inverter type have different impacts on the mode's stability. 
For example, most GFM inverters with P-f and Q-$V_\text{AC}$ droop controllers stabilise $\lambda_3$, while most GFM configurations with P-f droop control only, destabilise it. 
Moreover, within the same inverter control configuration, changes in parametrisation can lead to different impacts. 

Overall, these results show that the inverter selection cannot always be reduced to a simple choice between GFL and GFM. 
While in some cases, this is sufficient to avoid destabilisation, the inverter control configuration and parametrisation generally play an important role in determining the inverter's impact on system strength. 
This highlights the usefulness of the ICST in connection planning, as it assesses inverters not only by type but also by control configuration and parametrisation, enabling the selection of designs that enhance system strength.

The applicability of the ICST is validated by comparing its results with the behaviour expected from the known causality of the modes. 
For the GFL-induced modes $\lambda_1\!$ and $\lambda_2\!$ whose dominant participants are the GFL inverters at buses 19 and 20, and 32 and 33 respectively, a GFM connection is expected to stabilize these modes, while additional GFL inverters would destabilize them. 
This behaviour is reflected in the ICST results for $\lambda_1$ and $\lambda_2$ in \figurename \ref{fig:Screening3Modes} (a) and (b), respectively. 
In contrast, for $\lambda_ 3$, whose dominant participants are the GFM inverters at buses 27 and 29, the ICST results show that selecting an inverter type and configuration is less straightforward. 

\begin{figure*}[t]
  \centering
  % \vspace{0cm}
\includegraphics[width=0.25\linewidth,angle=-90]{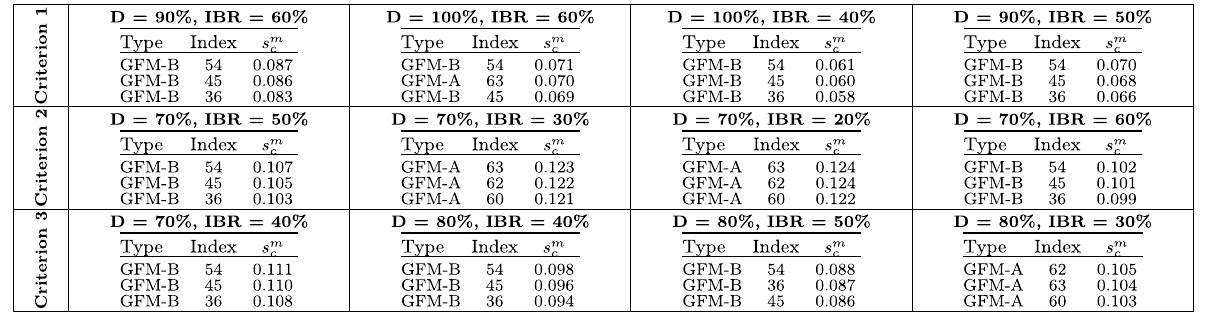}
  \caption{Top three scoring inverter configurations for each of the operating scenarios selected using the three COP criteria}
  % \vspace{-0.2cm}
  \label{fig:best_inverters}
\end{figure*}

Having evaluated the modal impact of each candidate configuration, the next step is to select the most suitable inverter configuration across all modes of interest and COPs.
Following the methodology in Section \ref{sub:inverter_selection} and using the weights in Table \ref{tab:ranges}, the suitability index of each inverter configuration \emph{m} at each COP \emph{c}, $s^\text{($m$,25)}_c$, is calculated. 
This calculation is repeated for all COPs identified for bus-25,  yielding an inverter suitability ranking for each COP. 
The three highest-ranked inverter configurations for each COP are shown in \figurename \ref{fig:best_inverters}, together with their suitability indices.
The maximin algorithm introduced in \eqref{eq:maximin_1} and \eqref{eq:maximin_2} is then applied to identify the best overall inverter configuration. 
The four highest-ranked inverter configurations across all COPs are listed in Table \ref{tab:maximin}. 

\begin{table}[t]
    \centering
        \caption{Final Inverter selection using maximin algorithm across COPs}
        \begin{tabular}{|c|c|c|c|c|}\hline\hline
        Rank & Type & Subtype Index (\emph{m}) &  $\underline{s}^\text{($m$,25)}$ &Worst-case COP\\\hline\hline
        \rowcolor[HTML]{c9dbbf}
        1&  GFM-B&  54& 0.0615 &D:100\%, IBR:40\%\\\hline
         2&  GFM-B&  45& 0.0603
 &D:100\%, IBR:40\%\\\hline
         3&  GFM-B&  36& 0.0583
 &D:100\%, IBR:40\%\\\hline
         4&  GFM-B&  53& 0.0477
 &D:100\%, IBR:40\%\\\hline
           % 5&  GFM-B&  44& 0.0467\\  \hline
         \hline
    \end{tabular}
      % \vspace{-0.2cm}
    \label{tab:maximin}
\end{table}
\begin{figure}[t]
    \centering
    \includegraphics[width=.9\linewidth]{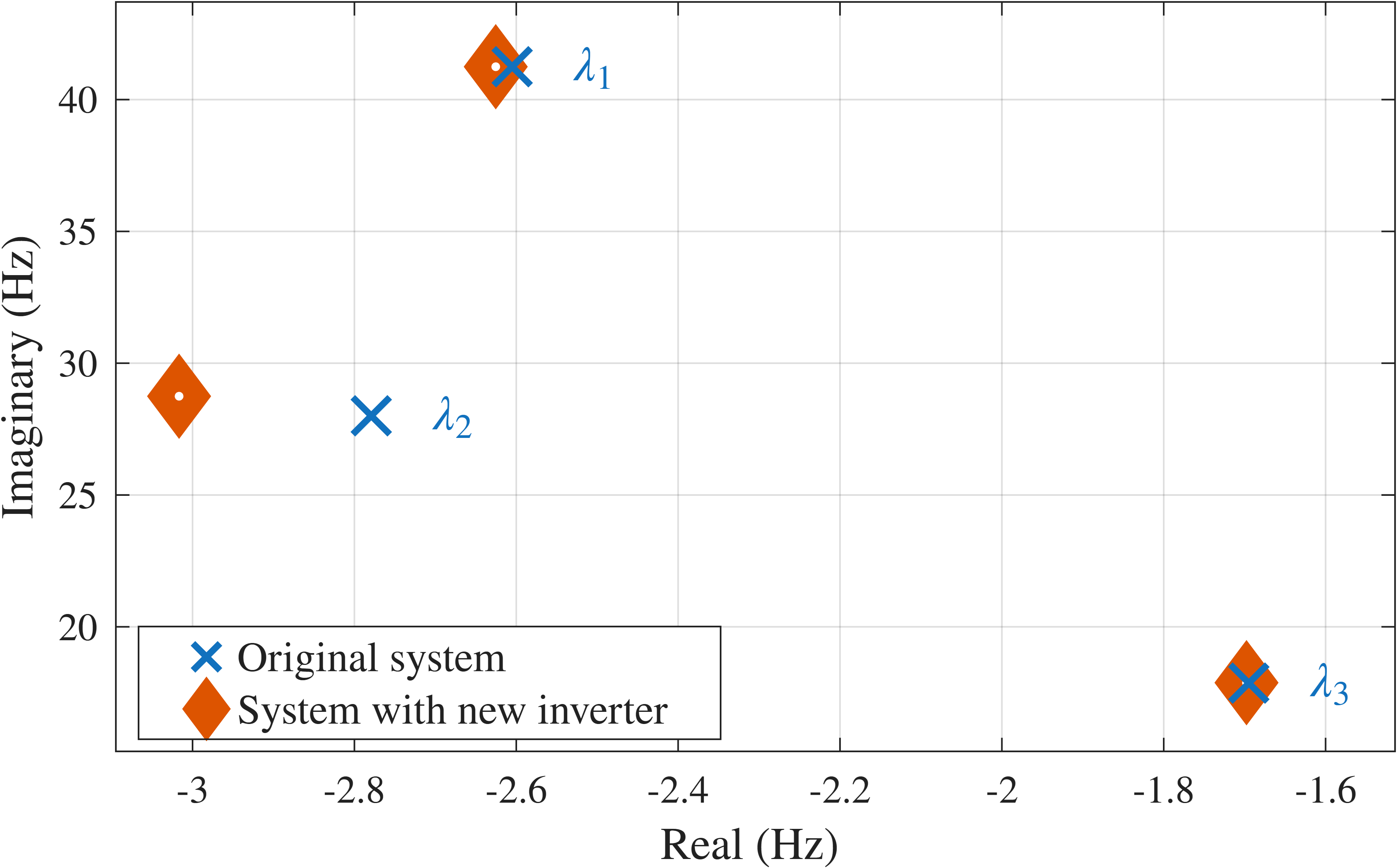}
     \caption{Modes of interest shift after the selected inverter connection }
    \label{fig:polemap_before_after}
      \vspace{-0.2cm}
\end{figure}
\begin{figure}[t]
    \centering
\includegraphics[width=0.9\linewidth]{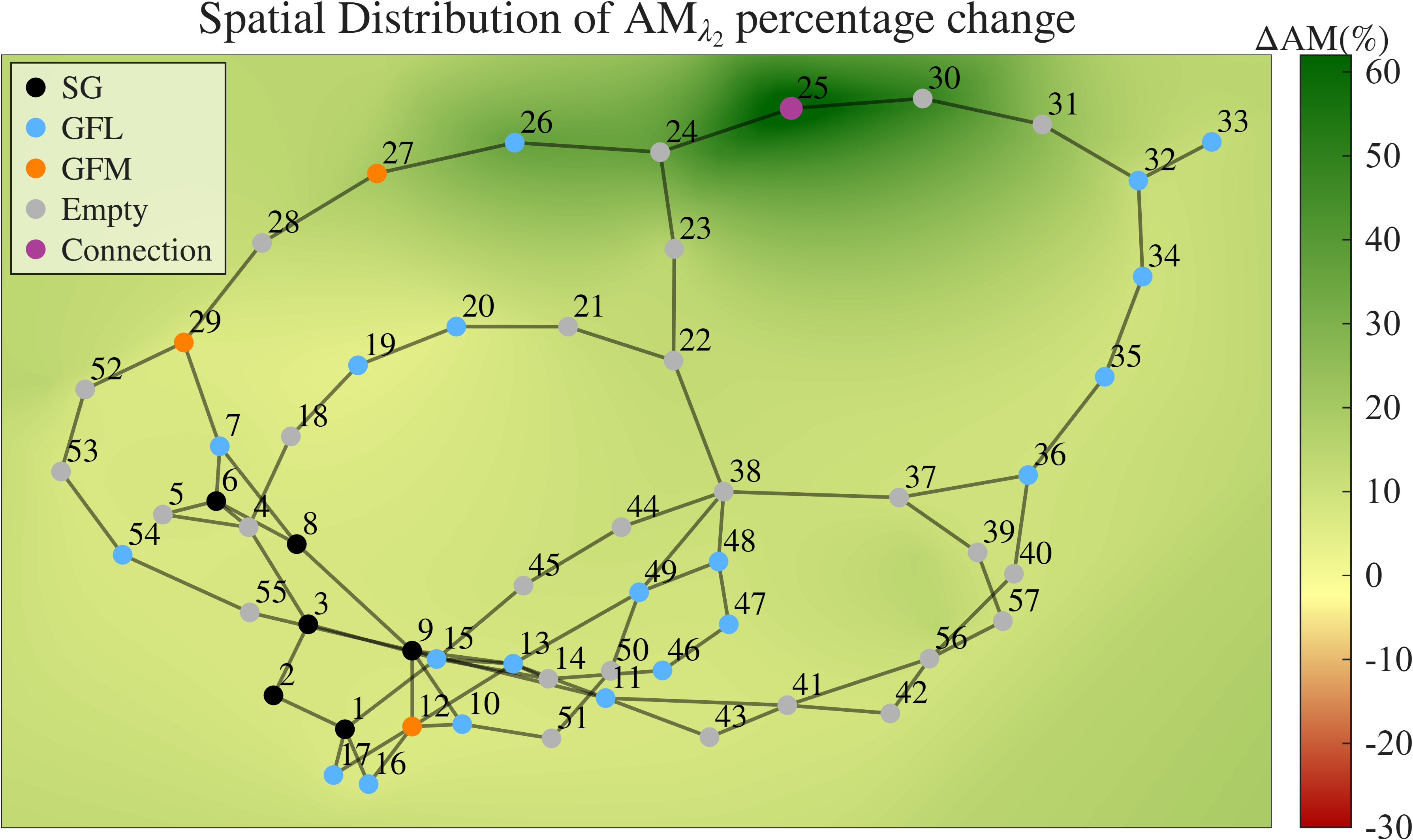}
    \caption{$\mathrm{AM}_{\lambda_2}$ percentage change heatmap upon selected inverter connection}
    \label{fig:am_change_heatmap}
\end{figure}
The GFM-B inverter configuration (GFM with P-f droop control) with parametrisation index 54, highlighted in Table \ref{tab:maximin}, is identified as the most suitable configuration for connection at bus-25. 
To verify this selection, the inverter is connected at bus-25, for the COP corresponding to its minimum suitability index, that is, Demand = 100\% and IBR = 40\%.
Since this is the worst-case COP for the selected inverter, this test provides a conservative validation of its impact. 
The whole-system impedance is then recalculated to determine the updated positions of the three modes of interest.
These are compared with the pre-connection mode positions in the pole map shown in \figurename\ref{fig:polemap_before_after}.
The results show that the connection improves the damping of $\lambda_2$, the most critical mode for bus-25, for which $\text{AM}_{\lambda_2}= 0.92$ under COP 100/40. 
The stabilising effect on $\lambda_1$ is considerably smaller, consistent with its higher $\text{AM}$ of 20.96, while its effect on $\lambda_3$ is negligible, as expected since $\lambda_3$ has the largest AM of the three modes, 52.03.
Finally, \figurename \ref{fig:am_change_heatmap} shows the spatial distribution of the percentage change in $\text{AM}_{\lambda_2}$ following the inverter connection at bus-25. 
A substantial increase in AM is evident across all buses, with the largest improvement occurring near the connection bus-25, further demonstrating the system-strength enhancement achieved by the selected inverter connection.

\vspace{-0.2cm}

\section{Conclusion}\label{conclusions}

This paper has investigated the challenges of planning new IBR connections from a small-signal system strength perspective and proposes an inverter connection screening tool (ICST) that can accurately evaluate candidate inverter configurations without requiring computationally intensive model-based studies. 
The ICST uses inverter admittances at critical modal frequencies and identifies the most suitable inverter configuration for a given connection location. 
It considers operating mode (GFL or GFM), control configuration, control parametrisation, and operating conditions, all of which were seen to strongly influence an inverter's impact on small-signal system strength. 
It has also been illustrated how the ICST can be applied by either a system operator or an IBR developer.
The accuracy of the ICST in predicting the impact of inverter connections on system modes has been validated against model-based eigenvalue analysis of the IEEE 57-bus system modified with 22 additional IBRs.
The case studies demonstrate its effectiveness in identifying inverter configurations that enhance system strength, mitigate the adverse effects of SSOs, and support higher levels of IBR integration. 
The results also show that the stabilising impact of an inverter connection is mode- and location-dependent, with a GFM inverter improving damping in some cases and a GFL inverter being more effective in others. 
Moreover, in certain cases, only particular control-loop configurations of a GFL or GFM inverter may be stabilising, while others may be destabilising.

% References section
\bibliographystyle{IEEEtran}
\bibliography{AMScreening}
 
% \section{Biography Section}
\vfill

\end{document}